\documentclass[preprint,12pt]{elsarticle}%
\usepackage{amssymb}
\usepackage{amsfonts}
\usepackage{amsmath}%
\usepackage{graphicx}
\providecommand{\U}[1]{\protect\rule{.1in}{.1in}}
\graphicspath{{E:/dottorato/Coulomb_damper/Testo/Paper/Immagini/}}
\journal{journal}

\begin{document}
%
%TCIMACRO{\TeXButton{Begin frontmatter}{\begin{frontmatter}}}%
%BeginExpansion
\begin{frontmatter}%
%EndExpansion



%TCIMACRO{\TeXButton{Title}{\title
%{A note on the possibility of roughness enhancement of adhesion in Persson's theory}%
%}}%
%BeginExpansion
\title
{A note on the possibility of roughness enhancement of adhesion in Persson's theory}%
%EndExpansion


%TCIMACRO{\TeXButton{Author}{\author{M. Ciavarella}}}%
%BeginExpansion
\author{M. Ciavarella}%
%EndExpansion
%

%TCIMACRO{\TeXButton{Address}{\address
%{Politecnico di BARI. Center of Excellence in Computational Mechanics. Viale Gentile 182, 70126 Bari. Mciava@poliba.it}%
%}}%
%BeginExpansion
\address
{Politecnico di BARI. Center of Excellence in Computational Mechanics. Viale Gentile 182, 70126 Bari. Mciava@poliba.it}%
%EndExpansion
%

%TCIMACRO{\TeXButton{Begin abstract}{\begin{abstract}}}%
%BeginExpansion
\begin{abstract}%
%EndExpansion

In an attempt to model the observed enhancement of adhesion in some classical
experiments in the 1970-1980's, Persson introduced in his theory of adhesion
between rough solids a term which corresponds to an area increase due to
roughness. In the old experiments, the adhesion enhancement was shown to be up
to one order of magnitude, whereas the area increase could not be defined
quantitatively because of possibly multiscale roughness. However, in more
recent studies by Guduru and collaborators, this enhancement has been further
explained with classical Linear Elastic Fracture Mechanics theory, the area
enhancement has been shown to be negligible, and therefore the problem of
adhesion of rough surfaces remains qualitatively and quantitatively unsolved
by Persson's theory.%

%TCIMACRO{\TeXButton{End abstract}{\end{abstract}}}%
%BeginExpansion
\end{abstract}%
%EndExpansion
%

%TCIMACRO{\TeXButton{Begin keyword(s)}{\begin{keyword}}}%
%BeginExpansion
%EndExpansion
%

%TCIMACRO{\TeXButton{End frontmatter}{\end{frontmatter}}}%
%BeginExpansion
\end{frontmatter}%
%EndExpansion



\section{\bigskip Introduction}

In his elegant theory of adhesion of rough surfaces, Persson (2002) (see also
Persson and Tosatti, 2001), in an attempt to justify some observations in
Briggs \& Briscoe (1977), and Fuller \& Roberts (1981), postulate that an
increase of adhesion may occur for the increase of surface area induced by
roughness. This is clearly stated in Persson and Tosatti (2001) \textit{"for
an (elastically) very soft solid the adhesion force may increase upon
roughening the substrate surface. This effect has been observed experimentally
[9]\footnote{Ref.9 is here Fuller \&\ Roberts (1981).}, and the present theory
explains under exactly what conditions that will occur}", and in Persson
(2002) "\textit{The increase in }$\Delta\gamma_{eff}$\textit{ arises from the
increase in the surface area}". Fuller \& Roberts (1981) show that the
adhesion enhancement can be up to one order of magnitude, and hence very
significant indeed, and classical asperity models (Fuller and Tabor, 1975)
were not able to capture this effect. However, in those experiments roughness
was random and possibly multiscale and not characterized fully, and hence a
good estimate of the area enhancement was not possible.

In general, Persson's theory "is valid for surfaces with arbitrary random
roughness", and in Sec. 5 they use it with success with the case of Fuller and
Tabor (1975) which correspond to an elastic sphere against a rough plate. In
the case of self-affine fractals, the power spectrum as a function of
wavevector $q$ has the form
\begin{equation}
C\left(  q\right)  =\left(
\begin{array}
[c]{cc}%
0 & \text{for }q<q_{0}\\
\frac{H}{2\pi}\left(  \frac{h_{0}}{\lambda_{0}}\right)  ^{2}\left(  \frac
{q}{q_{0}}\right)  ^{-2\left(  H+1\right)  } & \text{for }q>q_{0}%
\end{array}
\right)
\end{equation}
where $H=3-D_{f}$ (with $D_{f}$ being the fractal dimension of the surface
comprised between 2 and 3), and $q_{0}$ is the lower cut-off wavevector which
corresponds to the largest wavelength in the spectrum, and can be due to
macroscopic shape. 

They indeed obtain an "effective adhesion energy" which is
"magnification-dependent", where $\zeta=q/q_{0}$ is the magnification factor,
which shows two competitive factors: i) and \textit{enhancement} due to the
area increase due to roughness, which for surface gradient $\triangledown
h\left(  x\right)  <<1$
\begin{equation}
A=A_{0}+\frac{1}{2}\int\sqrt{1+\left(  \triangledown h\right)  ^{2}}dS\simeq
A_{0}+\frac{1}{2}\int\left(  1+\frac{1}{2}\left(  \triangledown h\right)
^{2}\right)  dS
\end{equation}
and ii) a decay due to the elastic deformation. This results in%
\begin{equation}
\Delta\gamma_{eff}=\Delta\gamma\left(  1+\left(  q_{0}h_{0}\right)
^{2}\left(  \frac{g\left(  H,\zeta\right)  }{2}-\frac{1}{q_{0}\delta}f\left(
H,\zeta\right)  \right)  \right)  \label{persson-result}%
\end{equation}
where $\delta=\frac{4\Delta\gamma}{E^{\ast}}$ , $E^{\ast}$ is plane strain
elastic modulus, and
\begin{align}
g\left(  H,\zeta\right)    & =\frac{H}{2\left(  1-H\right)  }\left(
\zeta^{2\left(  1-H\right)  }-1\right)  \rightarrow\zeta^{2\left(  1-H\right)
}\label{g}\\
f\left(  H,\zeta\right)    & =\frac{H}{1-2H}\left(  \zeta^{1-2H}-1\right)
\rightarrow\zeta^{1-2H}\label{f}%
\end{align}
The function $g$ is of the order of 100-500 in the original plots of Persson
and Tosatti (2001), and strongly depends on magnification $\zeta$, but clearly
these numbers do not have much sense, since the equation they use is obtained
in the limit $\triangledown h\left(  x\right)  <<1$, and at $\triangledown
h\left(  x\right)  =1$ obviously we only have a mere 41\% increment, which is
indeed the maximum enhancement which Persson shows in his FIG. 2 of Persson,
(2002). Beyond this point, we would be in areas where finite deformations, and
many other deviations from the usual approximations would happen. Notice that
the function $f$ decreases if $H>0.5$ (fractal dimension $D<2.5$, which is the
common case), indicating that the effective adhesion energy tends to return to
the original value without roughness --- a result that is however not as clear
as this analysis is limited by the strong assumption of full contact which is
so far uncontrolled.

Even when both functions grow, $g$ grows much faster than $f$, and the authors
do not suggest where we should stop. However, the point is not this, but that
the area increase is completely unrelated to the adhesion enhancement of
Briggs \& Briscoe (1977), and Fuller \& Roberts (1981).

\bigskip Indeed, after Persson's theory has appeared, in very interesting
experiments using a single scale axisymmetric roughness between gelatin and
Perspex flat rough plates, by Guduru and his group (Guduru (2007), Guduru
\&\ Bull (2007), Waters \textit{et al} (2009)), the adhesion enhancement has
been studied in details, and shown to be of an order of magnitude even when
the surface area increase is (as we easily estimate below) of much less than
0.1\%. 

\bigskip Waters \textit{et al} (2009) have a good summary of Guduru's group
theory and experiments. They have a surface defined as%
\begin{equation}
f\left(  r\right)  =\frac{r^{2}}{2R}+A\left(  1-\cos\frac{2\pi r}{\lambda
}\right)
\end{equation}
where $A$ can be both positive in the case of a central convex asperity, and
negative, for a central concave trough. 

The enhancement is shown to occur when complete contact occurs and the contact
area is simply connected. In the earlier paper (Guduru (2007)), conditions
were derived for the gap to be monotonically increasing with radius, but this
condition is overly restrictive, as it is well known even from Persson's
energy balance concept, that adhesion permits a wavy surface to spontaneously
achieve full contact. The analysis follows conveniently introducing two
parameters%
\begin{equation}
\alpha=\frac{AR}{\lambda^{2}}\quad,\quad\beta=\frac{\lambda^{3}E^{\ast}}%
{2\pi\Delta\gamma R^{2}}%
\end{equation}
The physical meaning of $\alpha$ is that obviously it represents the degree of
surface waviness. There are only two scales really in the process, one
represented by the radius of the sphere $R$ (no specific reference to the
amplitude and waviness), and the other by amplitude and wavelength of
roughness: $\alpha$ is also obviously the ratio between the radii of the
sphere and that of the asperities. Large $\alpha$ correspond to surfaces with
high amplitude, short wavelength waviness. The parameter $\beta$ is instead a
measure of the relative stiffness of the material to the surface energy.\ The
adhesion amplification is seen in a clear map in Fig.5 of \bigskip Waters
\textit{et al} (2009) for the JKR regime. It is seen to reach values over 4
(in terms of pull-off, but equivalently in terms of $\Delta\gamma_{eff}$ in
Persson's theory notation), for values of $\alpha<0.25.$

Guduru and Bull (2007) demonstrated the actual validity of these predictions,
with experiments with soft gelatin, with waviness amplitude $A=1.2\mu m$ and
wavelength $\lambda=0.2mm$. This corresponds to an estimate increase of area
of $1+\left(  \frac{\pi}{2}\frac{A}{\lambda}\right)  ^{2}=1+\left(  \frac{\pi
}{2}\frac{1.2}{0.2}10^{-3}\right)  ^{2}\simeq1.0001$, whereas the pull-off
increase was a factor about 2. Even worse the comparison with waviness
amplitude $A=5.5\mu m$ and wavelength $\lambda=0.43mm$. Here, $1+\left(
\frac{\pi}{2}\frac{5.5}{0.43}10^{-3}\right)  ^{2}=\allowbreak1.\,\allowbreak
000\,4$, while the amplification factor was about 6. These examples illustrate
the very different nature of increases in adhesive strength resulting from the
presence of shallow waviness on soft elastic surfaces.

Notice that these results occur in a situation where the large amplitude of
roughness is in partial contact, and the roughness scale is in a full contact.
Persson's theory takes into account of the possibility of partial contact in
later parts of the papers, but only to derive further \textit{reduction} of
adhesion, and certainly not increase. Hence, the partial contact correction
can only make the comparison worse.  \bigskip In Fig. 5 of Guduru and Bull
(2007), it is shown that beyond a critical $\beta$, there is a region where
the "enhanced" strength occurs only if the contact is loaded first
sufficiently to cause full contact in the roughness scale. Beyond an even
greater $\beta$, finally a reduction in pull-off force for the wavy surface
compared to the flat surface occurs, when locally the contact is one of two
isolated spherical asperities with a much reduced equivalent radius. However,
even in this range, an increase of $\alpha$ leads to an increase of adhesion.
Therefore, not even this is the regime usually indicated in asperity models
like Fuller and Tabor (1975) as roughness destroying adhesion. There are other
aspects of Guduru's enhancement which are not considered in Persson's theory,
and a remarkable one is the absence of irreversible processes leading to an
increase of toughness. There is a qualitative discussion at the end of par.4
of Persson and Tosatti (2001) about this aspect, but none of these effects is
included in the theory.

\section{Discussion}

The Guduru enhancement of adhesion could be even stronger for multiscale
roughness, and the limitations will be (i) that some adhesion enhancements
will be load-dependent; and (ii) that Guduru's analysis considers separation
originating at the periphery, which may be increasingly a strong assumption
when multiscale roughness is included. We have recently attempted to consider
the possibility of separation to occur at the local minima of the surface
waves, where tensile interface stresses will be highest for a Gaussian random
roughness (Ciavarella, 2016). The analysis shows that a very simple
approximate solution is possible: we consider the full contact solution which
is known in closed form, and consider the condition for the gaps in regions of
tensile stresses to remain open or close. This leads to a solution very
similar to Persson's solution in contact mechanics without adhesion (Persson,
2001), namely that
\begin{equation}
\frac{A_{c}\left(  0\right)  }{A_{0}}=\operatorname{erf}\left(  \frac{\sqrt
{2}}{E^{\ast}}\frac{\left(  \overline{p}+p_{\min}\right)  }{\sqrt{V}}\right)
\end{equation}
which is valid for positive mean pressure $\overline{p}>0$ only, and where $V$
is the variance of full contact pressure variations. Also,
\begin{equation}
\frac{p_{\min}}{\sqrt{V}}\sim\zeta^{2/5\left(  2H-1\right)  }\label{p0min}%
\end{equation}
and therefore increases without limit with magnification for $H>0.5$ or for
fractal dimension of the surface $D<2.5$. It wasn't noticed in (Ciavarella,
2016) that this model show therefore that under zero load the ratio of the
area of contact to the nominal area would be%
\begin{equation}
\frac{A_{c}\left(  0\right)  }{A_{0}}=\operatorname{erf}\left(  \frac{\sqrt
{2}}{2E^{\ast}}\frac{p_{\min}}{\sqrt{V}}\right)
\end{equation}
and therefore a propensity of reaching full contact, but never obviously the
exact full contact. However, notice the analogy with Persson's function $f$
which decreases if $H>0.5$. Here, we obtain this result about the contact area
directly considering partial contact, and only with some small approximations
to obtain the closed form results (for details, see Ciavarella, 2016). We do
obtain that for $D<2.5$, the common case, the contact area tends to be
complete. We would be tempted to say that this seems to give some meaning of
"effective adhesion energy": here it takes the sense of the energy available
when we start the process of unloading, but very little can be said about the
reversible and irreversible processes that start upon unloading, nor the
maxima we could reach of pull-off force. Persson seems to take another more
meaning, of effective energy assuming full contact, which has to be corrected
considering partial contact, resulting in a circular definition. 

The situation of unloading unfortunately cannot be treated with this model, as
cannot be treated in Persson's theory of adhesion, and therefore, the problem
remains largely unsolved. 

Finally, another limitation of the Guduru effect will be at small scales, in
that the Waters \textit{et al} (2009) show the enhancement to be limited to
the JKR regime (Johnson \textit{et al.}, 1971), whereas the small scales are
essentially in the DMT regime.

\section{Conclusions}

The large increments of adhesion measured in soft solids cannot be captured by
Persson's model of adhesion. Therefore, the competition between these adhesion
enhancement with multiscale roughness has not yet been understood.

\section{References}

Briggs G A D and Briscoe B J (1977) The effect of surface topography on the
adhesion of elastic solids J. Phys. D: Appl. Phys. 10 2453--2466

Ciavarella, M. (2016). Adhesive rough contacts near complete contact. in
press, Int J Mech Sci., arXiv preprint arXiv:1504.08240.

Fuller, K. N. G., \& Tabor, D. (1975). The effect of surface roughness on the
adhesion of elastic solids. Proc Roy Soc London A: 345, No. 1642, 327-342

Fuller, K.N.G. , Roberts A.D. (1981). Rubber rolling on rough surfaces J.
Phys. D Appl. Phys., 14, pp. 221--239

Guduru, P.R. (2007). Detachment of a rigid solid from an elastic wavy surface:
theory J. Mech. Phys. Solids, 55, 473--488

Guduru, P.R. , Bull, C. (2007). Detachment of a rigid solid from an elastic
wavy surface: experiments J. Mech. Phys. Solids, 55, 473--488

Johnson, K. L., K. Kendall, and A. D. Roberts. (1971). Surface energy and the
contact of elastic solids. Proc Royal Soc London A: 324. 1558.

Persson, B.N.J., (2001). Theory of rubber friction and contact mechanics. J.
Chem. Phys. 115, 3840--3861.

Persson, B.N.J. (2002). Adhesion between an elastic body and a randomly rough
hard surface, Eur. Phys. J. E 8, 385--401

Persson, B. N. J., \& Tosatti, E. (2001). The effect of surface roughness on
the adhesion of elastic solids. The Journal of Chemical Physics, 115(12), 5597-5610.

Waters, J.F. Leeb, S. Guduru, P.R. (2009). Mechanics of axisymmetric wavy
surface adhesion: JKR--DMT transition solution, Int J of Solids and Struct 46
5, 1033--1042

\end{document}